\documentclass[abbreviations]{article}
 
\usepackage{amssymb,amsmath,amsthm} 
\usepackage{graphicx}
\usepackage[latin1]{inputenc}
\usepackage{acronym}
\usepackage[a4paper, centering, text={6.5in, 10in}]{geometry}
\usepackage{authblk}

\newtheorem{theorem}{Theorem}[section]

\newtheorem{definition}[theorem]{Definition}
\newtheorem{example}{Example}[section]
\newtheorem{remark}[example]{Remark}

\providecommand{\keywords}[1]{\textbf{\textit{Index terms---}} #1}

\author{Francisco Mota}
\affil{Departamento de Engenharia de Computação e Automação\\
Universidade Federal do Rio Grande do Norte -- Brasil\\
e-mail:mota@dca.ufrn.br}

\date{\today}

\title{Projective Root-Locus: An Extension of Root-Locus Plot to the
Projective Plane}

\begin{document}

\maketitle

\begin{abstract}
In this paper we present an extension of the classical Root-Locus 
(RL) method  where the points are calculated in the real projective plane 
instead of the conventional affine real plane; we denominate this 
extension of the Root-Locus as ``Projective Root-Locus (PjRL)". 
To plot the PjRL we use the concept of ``Gnomonic Projection'' in order 
to have a representation of the projective real plane as a semi-sphere of 
radius one in ${\mathbb R}^3$. We will see that the PjRL reduces to the RL 
in the affine $XY$ plane, but also 
we can plot the RL onto another affine component of the projective 
plane, like $ZY$ affine plane for instance, to obtain what we denominate 
complementary plots of the conventional RL. We also 
show that with the PjRL the points at infinity of the RL can be 
computed as solutions of a set algebraic equations.


\end{abstract}

\section{Introduction}

\begin{figure}
\begin{center}
\includegraphics{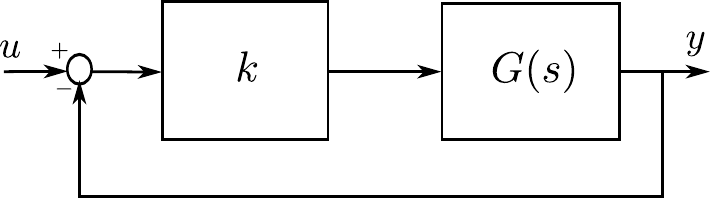}
\caption{\label{cloop}Control Feedback Loop with a Proportional Controller}
\end{center}
\end{figure}

The Root-Locus (RL) method is a classical tool that has been used
extensively in the feedback control literature
for studying the stability and performance of a closed loop linear
feedback system.
It consists of a parametric plot of the roots of the polynomial
$p(s) = d(s) + kn(s)$ in in the complex plane, 
as the parameter $k$ spans $\mathbb{R}$; $d$ and $n$
are fixed coprime polynomials, and $d$ is monic with degree, in general, greater
than the degree of $n$. In fact, the polynomial $p$
represents the denominator of the transfer function of a closed-loop feedback
system that has the (irreducible) proper rational function $G(s) = n(s)/d(s)$ as a
linear time invariant plant model and $k$ as a (proportional type) controller (see
Figure~\ref{cloop}) and that is why we use the terminology the
``RL for $G(s)$''. To plot the RL for a given $G(s)$, most control theory
textbooks presents a set
of rules that allow us to make an approximate sketch of the plot (\cite {dh}), but a
detailed plot, nowadays, in general, is obtained using a computer software
that evaluates the roots of $p$, using numerical
techniques, for a given range of the parameter $k$ in $\mathbb{R}$
(e.g. Scilab (\cite{scilab})). In
Figure~\ref{rl1} we show the plot of the RL for the plant 
$G(s) = (s+1)/s^2$, for some range of $k\in\mathbb R$. 

The motivation to use the projective plane to analyze the RL
method is that the RL plot for a given
$G(s)$ can have points at infinity: the parameter $k$ itself has
to reach an ``infinite value'' in order we can obtain the ``terminal''
points of the RL, that can, in turn, be finite (zeros
of $G(s)$) or to be located at infinity. In this way, using the concepts
of projective real line and projective plane 
we can account for these ``infinite'' points, and also obtain complementary
plots of the RL where points at infinity can be plotted at a 
finite position onto an affine plane.
We denominate this extension of the
RL to the projective plane as  ``Projective Root-locus (PjRL)'' and, 
in spite of its abstract definition, we will show that it can be 
relatively easy to obtain the PjRL for $G(s)$
using a computer algebra software. Below we
introduce the definitions and notation to be used along the paper:

\begin{figure}
\begin{center}
\includegraphics[scale=1.5]{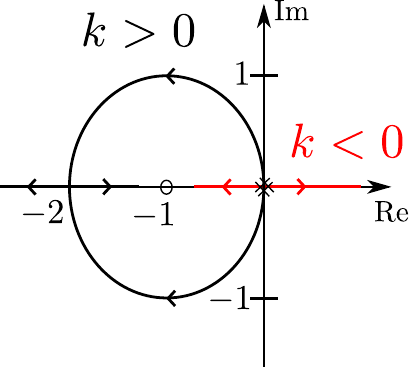} 
\caption{\label{rl1} Root-Locus for $G(s)=(s+1)/s^2$} 
\end{center}
\end{figure} 

\begin{description}

\item[$\pmb{\mathbb R}$, $\pmb{\mathbb C}$] {\bf and} $\pmb{\mathbb R[x_1, x_2, \ldots, x_n]}$:
Represents the field of real numbers, the
field of complex numbers and the ring of polynomials with coefficient's in
$\mathbb R$ and with indeterminates 
$(x_1, x_2, \ldots, x_n)$, respectively.

\item[Projective (real) line:] The projective line
$\mathbb{P}^1(\mathbb{R})$ is the set of ``slopes'' $y/x$,
$(x,y)\neq (0,0) \in \mathbb{R}^2$ and $1/0=\infty$. So, if
$k\in\mathbb{P}^1(\mathbb{R})$, then
$k=k_n/k_d$, and $k=\infty$ corresponds to $k_n=1$ and $k_d=0$
($k_n=k_d=0$ is not allowed).
We note that $\mathbb{P}^1(\mathbb{R})=\mathbb{R}\cup \{\infty\}$.

\item[Projective (real) plane:] The projective plane
$\mathbb{P}^2(\mathbb{R})$ is the set of equivalence classes of all
{\em nonzero} triples
$(x,y,z)\in \mathbb{R}^3 $ under the equivalence relation:
$(\alpha_1,\alpha_2,\alpha_3) \equiv (\beta_1,\beta_2,\beta_3)$
if $\alpha_i=\lambda \beta_i$, for some $\lambda \neq 0$. We represent the
equivalence class of $(x,y,z)$ by $(x:y:z)$, that is
denominated ``homogeneous coordinate'' of $(x,y,z)$. We
note that $\mathbb{P}^2(\mathbb{R}) = \mathbb{R}^2 \cup H$, where $H$
represents the plane at infinity and it is disjoint from
$\mathbb{R}^2$. Mathematically, $\mathbb{R}^2 = \{(x:y:1) \in
\mathbb{P}^2(\mathbb{R})\}$, the $XY$ plane, and
$H = \{(x:y:0) \in \mathbb{P}^2(\mathbb{R})\}$. In fact, $H$ has two
``types'' of points, namely, $(1:m:0)$ and
($0:1:0)$, where $(1:m:0)$ represents the point of intersection of all
($XY$) lines  with finite slope ``$m$'' and
$(0:1:0)$ represents the intersection of all lines with infinite slope
(vertical lines).

\item [Homogeneous polynomial] A polynomial (in several variables) is
homogeneous when all of its nonzero terms (monomials)
have the same total degree. 
One important fact about a homogeneous polynomial
$p\in\mathbb R[x_1, x_2, \ldots, x_n]$ is that
$p(\lambda x_1, \lambda x_2, \ldots, \lambda x_n) = \lambda^d p(x_1, x_2,
\ldots, x_n)$, where $d$ is the total degree of $p$. We always
can turn a non-homogeneous polynomial ($q$) into a homogeneous one ($q^h$)
by adding a new variable ($x_{n+1}$), with the following procedure:
$q^h(x_1, \ldots, x_n,x_{n+1}) = x_{n+1}^{d}\;q(x_1/x_{n+1}, x_2/x_{n+1}, 
\ldots, x_n/x_{n+1})$, where $d$ is the total degree of $q$; this process is
denominated ``homogenization'' of $q$.   We can always ``de-homogenize"
$q^h$ by setting $x_{n+1}=1$ and recover back $q$.

\item[Affine Algebraic Variety] An affine (real) algebraic variety
${\cal V}$ generated by a set of $m$ polynomials,
$p_i\in \mathbb R[x_1,x_2,\ldots, x_n]$, is a subset of the affine plane
$\mathbb R^n$ composed by the coordinates
$(x_1,x_2, \ldots, x_n)$ that are simultaneous {\em real} roots of the $m$ generating
polynomials, that is  $p_i(x_1, x_2, \ldots, x_n)=0$, $i=1,\ldots, m$.

\item[Projective Algebraic Variety] A projective (real) algebraic variety
${\cal W}$ generated by a set of $m$
{\em homogeneous} polynomials, $p_i\in \mathbb R[x_1,x_2,\ldots,
x_n, x_{n+1}]$, is a subset of the projective plane
$\mathbb P^n(\mathbb R)$ composed by the
{\em homogeneous} coordinates $(x_1:x_2: \ldots :x_n:x_{n+1})$ 
such that $(x_1,x_2,\ldots, x_n,x_{n+1})$ is a 
simultaneous {\em real} root of the $m$ generating
homogeneous polynomials, that is  $p_i(x_1, x_2, \ldots, x_n,x_{n+1})=0$,
$i=1,\ldots, m$. We note that if
$(x_1', \ldots, x_n',x_{n+1}')$ is any member of the equivalence class
$(x_1:x_2: \ldots :x_n:x_{n+1})$, it is also a simultaneous
root of $p_i$, $i=1,\ldots, m$, since $p_i$ is homogeneous of degree $d$: 
$x_j'= \lambda x_j$, then
$p_i(x_1', \ldots, x_n',x_{n+1}') = \lambda^d p_i(x_1, x_2, \ldots, x_n,x_{n+1})=0$.

\item[Ideal of Polynomials:] Let be $\{p_1, p_2, \ldots, p_t\}$ a set of
polynomials in $\mathbb R[x_1, x_2, \ldots, x_n]$.
The set of polynomials $I \subseteq  \mathbb R[x_1, x_2, \ldots, x_n]$
defined by
\[
I = \sum_{i=1}^{t}h_ip_i, \quad h_i \in \mathbb R[x_1, x_2, \ldots, x_n]
\]
is an ideal of $\mathbb R[x_1, x_2, \ldots, x_n]$, and 
$\{p_1, p_2, \ldots, p_t\}$ is denominated a {\em generating set for $I$}; in this case
we write $I=\langle p_1, p_2, \ldots, p_t\rangle$. A Grobner Basis for the ideal 
$I$ is a particular kind of generating set that allows many important 
properties of the ideal to be deduced easily. Given a generating 
set $\{p_1, p_2, \ldots, p_t\}$ for $I$, we can obtain a Grobner basis
$\{g_1, g_2, \ldots, g_s\}$ for $I$ algorithmically (see \cite[Ch.~2]{clo}).

\end{description}
For more details about the concepts above see (\cite{clo}, \cite{sha}).

\section{The Projective Root-Locus - PjRL}

As discussed in Introduction, the conventional RL for an irreducible 
proper rational function $G(s)=n(s)/d(s)$
is a plot of the roots of the polynomial 
$p(s) = d(s) + kn(s)$, when $k\in\mathbb{R}$; that is, we solve
\begin{equation}\label{pencil}
d(s) + kn(s) = 0
\end{equation}
for each $k\in\mathbb{R}$ and plot its roots in the affine
plane $\mathbb R^2$. But, since the 
parameter $k$ belongs to $\mathbb R$, to analyze the situation where $k\to \pm\infty$,
we will modify Equation~(\ref{pencil}) slightly by considering 
$k\in\mathbb{P}^1(\mathbb{R})$. So, following the definition of $\mathbb
P^1(\mathbb R)$, we set $k=k_n/k_d$ in (\ref{pencil})
and clear the denominator to obtain:
\begin{equation}\label{ppencil}
k_dd(s)+k_nn(s) = 0.
\end{equation}
We note that setting $k_d=1$ in Equation~(\ref{ppencil}) we recover 
Equation~(\ref{pencil}) and setting $k_d=0$, that is 
$k=\infty$, corresponds
to $n(s)= 0$ in Equation~(\ref{ppencil}), or the finite ``terminal'' points of
the RL (zeros of $G(s)$).  We then see that the effect of passing 
from $k\in\mathbb R$ to $k\in\mathbb P^{1}(\mathbb R)$
is just that of including the roots of $n(s)$, the finite terminal
points, into the RL. As we will see in the next sections, 
the ``infinite'' terminal points of the RL will only appear when we 
extrapolate from  $\mathbb R^2$ to $\mathbb P^2(\mathbb R)$.
We also note that we can treat the case where the 
degree of $d$ is less than the degree of $n$ 
in the same fashion we treat the case where the degree
of of $d$ is greater than the degree of $n$ by
just exchanging the positions of $k_d$ and $k_n$ in 
Equation~\ref{ppencil}. The case 
where the degree of $d$ is equal the degree of $n$ 
also can be treated by our approach, 
as shown in examples of Section~\ref{examples}.

Since Equation~(\ref{ppencil}) may admit complex solutions, 
if we write $s=x+iy$ we have:
\begin{equation}\label{qdrd}
d(x+iy) = q_d(x,y) + ir_d(x,y) \quad{\rm and}\quad n(x+iy) = q_n(x,y)+ir_n(x,y)
\end{equation}
where $q_d, r_d, q_n$ and $r_n$ are polynomials in 
$\mathbb R[x,y]$. So we may rewrite (\ref{ppencil}) as:
\[
\left[k_dq_d(x,y)+k_nq_n(x,y)\right] + 
i\left[k_dr_d(x,y)+k_nr_n(x,y)\right]=0
\]
and finding a complex solution for (\ref{ppencil}), for 
given pair $(k_d,k_n)$,  is 
equivalent of finding a solution in $\mathbb R^2$
for the system of polynomial equations:
\begin{eqnarray}
k_dq_d(x,y)+k_nq_n(x,y) & = & 0\label{rpencil1}\\
k_dr_d(x,y)+k_nr_n(x,y) & = & 0 \label{rpencil2}
\end{eqnarray}
for each $k_n/k_d\in \mathbb P^1(\mathbb R)$.   
It is important to stress the fact that  any solution for the 
system (\ref{rpencil1}--\ref{rpencil2}) must be invariant
when we pass from pair $(k_n,k_d)$ to $(\lambda k_n,\lambda k_d)$, $\lambda
\neq 0$, since they represent the same point in $\mathbb P^1(\mathbb R)$.
This, in fact, is true because it is equivalent to multiply
Equations (\ref{rpencil1}) and (\ref{rpencil2}) by $\lambda\neq 0$. 

To obtain the Projective Root-Locus (PjRL) 
we need to extend the solutions of 
Equations~(\ref{rpencil1}--\ref{rpencil2}), defined above, from
the affine plane $\mathbb R^2$, to the
projective plane $\mathbb P^2(\mathbb R)$.
To achieve this goal, we first need to interpret 
the solutions of Equation~(\ref{rpencil1}--\ref{rpencil2}) as 
a real algebraic variety ${\cal V}$ generated by the set of two polynomials 
$q$ and $r$ defined as:
\begin{eqnarray}
q(x,y,k_d,k_n) & = & k_dq_d(x,y)+k_nq_n(x,y)\label{poly1}\\
r(x,y,k_d,k_n)  & = & k_dr_d(x,y)+k_nr_n(x,y).\label{poly2}
\end{eqnarray}
Since the polynomials $q$ and $r$ are defined in 
$\mathbb R[x,y,k_d,k_n]$ we would have a 
variety in $\mathbb R^4$, i.e.\ ${\cal V}\subset \mathbb R^4$;
but, since $k_n/k_d$ is defined in $\mathbb P^1(\mathbb R)$,
in fact, we have 
${\cal V}\subset \mathbb R^2\times\mathbb P^1(\mathbb R)$.
Based on this, we could abstractly interpret the RL as the 
projection (represented by ${\cal V}_k$) of ${\cal V}$ onto $\mathbb R^2$,
since each point of the RL is a solution of (\ref{rpencil1}--\ref{rpencil2})
for a fixed $k=k_n/k_d\in\mathbb P^1(\mathbb R)$. We note that,
for each $k\in\mathbb P^1(\mathbb R)$, ${\cal V}_k$ is an 
(finite) affine real variety defined in $\mathbb R^2$, by the 
solutions of Equations (\ref{rpencil1}--\ref{rpencil2}),
or equivalently, by the roots of Equation~(\ref{ppencil}).

Now we proceed with the question of extrapolating the RL from the 
affine plane ($\mathbb R^2$) to the projective 
plane ($\mathbb P^{2}(\mathbb R)$). Our approach will
follow the two steps bellow:
\begin{enumerate}
\item[(1)] Extrapolate the algebraic variety 
${\cal V}\subset\mathbb R^2\times\mathbb P^{1}(\mathbb R)$,
defined above,
to obtain a projective algebraic variety 
${\cal W}\subset\mathbb P^2(\mathbb R)\times\mathbb P^{1}(\mathbb R)$;

\item[(2)] Obtain the projection  of 
${\cal W}$ onto $\mathbb P^2(\mathbb R)$. This projection, 
represented by ${\cal W}_k$, $k\in\mathbb P^1(\mathbb R)$,
it will be what we denominate PjRL.

\end{enumerate}

To obtain ${\cal W}$ from ${\cal V}$, we could
simply homogenize the 
polynomials $q$ and $r$, presented in Equations~(\ref{poly1}--\ref{poly2}),
and obtain a projective
variety ${\cal W}$ in $\mathbb P^2(\mathbb R)\times\mathbb P^{1}(\mathbb R)$, 
now generated by the homogenized polynomials 
$q^{h}(x,y,z,k_d,k_n)=z^dq(x/z,y/z,k_d/z,k_n/z)$ and 
$r^{h}(x,y,z,k_d,k_n)=z^er(x/z,y/z,k_d/z,k_n/z)$, 
as defined in Introduction. The projective variety obtained this way
reduces to ${\cal V}$ in $\mathbb R^2\times\mathbb P^1(\mathbb R)$,
since the process of de-homogenization of $q^{h}$ and $r^{h}$ will
restore back the polynomials $q$ and $r$. The flaw with
this approach is that the process of simply homogenizing the generating
polynomials for ${\cal V}$,
in general, creates a projective variety that is ``greater'' than the
necessary, in the sense that it may add
points at infinity to the original variety, 
other than the existing ones (see \cite[Ch.~8]{clo}).
Then, in fact, ${\cal W}$ must be the ``projective closure'' of ${\cal V}$,
that is, a {\em minimal} projective
variety in $\mathbb P^{2}(\mathbb R)\times\mathbb P^1(\mathbb R)$ 
that reduces to ${\cal V}$
in $\mathbb R^2\times\mathbb P^1(\mathbb R)$. 
To compute the closure of ${\cal V}$, 
instead of directly homogenizing the
polynomials $q$ and $r$ that generates ${\cal V}$, we need
first to compute a Grobner basis, with respect a
{\em graded} monomial order, for
the ideal $I=\langle q,r\rangle$ (see \cite[Ch.~8]{clo}). 
The projective closure of ${\cal V}$ will be the
projective variety ${\cal W}$ generated by the homogenized polynomials
of the obtained Grobner basis.
For the sake of completeness we present the following
definition for the PjRL:

\begin{definition}\label{pjrlc}\em
{\bf (PjRL)} Let be an irreducible rational function $G(s)=n(s)/d(s)$ and consider the
polynomials $q$ and $r$ as defined in Equations~(\ref{poly1}--\ref{poly2}).
We call the PjRL of $G(s)$ the projection onto $\mathbb P^2(\mathbb R)$ 
of the projective algebraic variety
${\cal W}\subset\mathbb P^2(\mathbb R)\times\mathbb P^{1}(\mathbb R)$, 
where ${\cal W}$ is generated by the set of  homogenized polynomials of the
Grobner basis $\{g_{1}, g_{2},\ldots, g_{s}\}$, with respect a graded monomial order, 
for the ideal $\langle q,r\rangle$. We will denote this projection 
by ${\cal W}_k$, where $k\in\mathbb P^1(\mathbb R)$. \qed
\end{definition}

\begin{remark}\em
We will denote the set homogenized polynomials of the Grobner basis for 
$\langle q,r\rangle$ by $\{g_{1}^{h}, \ldots, g_{s}^{h}\}$, where 
$g_{i}^{h}\in \mathbb R[x,y,z,k_d,k_n]$,  and the variable $z$ comes 
from the homogenization process, as defined in Introduction. Since
we analyze ${\cal W}_k$ in $\mathbb P^2(\mathbb R)$, we consider 
$k=k_n/k_d \in \mathbb P^1(\mathbb R)$ as a parameter, 
and the homogeneous polynomials 
$g_{i}^{h}$ can been seen as defined in $\mathbb R[x,y,z]$.
Based on the fact that $k\in \mathbb P^1(\mathbb R)=\mathbb R\cup \{\infty\}$, 
we define:
\begin{description}
\item[Initial points of the PjRL ($\pmb{{\cal W}_0)}$:] $k=0/1 = 0$; that is, ${\cal W}_0$ is generated 
by the polynomials $\{g_{1}^{h}, g_{2}^{h}, \ldots, g_{s}^{h}\}$  setting $k_n=0$ and $k_d=1$. 

\item[Terminal points of the PjRL ($\pmb{{\cal W}_{\infty}}$):] $k = 1/0=\infty$; that is, ${\cal W}_{\infty}$ 
is generated by the polynomials $\{g_{1}^{h}, g_{2}^{h}, \ldots, g_{s}^{h}\}$ setting $k_n=1$ and $k_d=0$. 

\item[Intermediary points of the PjRL ($\pmb{{\cal W}_{\lambda}}$):] $k = \lambda/1, \lambda\neq 0$;
that is  ${\cal W}_{\lambda}$  is generated by the polynomials $\{g_{1}^{h}, g_{2}^{h}, \ldots, g_{s}^{h}\}$ 
setting $k_n=\lambda\in\mathbb R\backslash\{0\}$ and $k_d=1$.  
\end{description} \qed
\end{remark}

We have the following comments regarding the results presented above:

\begin{itemize}

\item Calculating the Grobner basis 
for the ideal $\langle q,r\rangle$ is a relatively 
easy procedure using an algebra software available such as Macaulay2 (\cite{mac}), 
since we have only two polynomials that 
depends on four indeterminates, namely $x,y,k_d$ and $k_n$. 

\item In the classical RL method there is a procedure for calculating the 
asymptotes based on the difference between the number of poles and zeros 
of $G(s)$. In our case, the direction of these 
asymptotes will appear as the solution of the 
algebraic equations that defines ${\cal W}_{k}$ 
and it will represent points at 
infinity. 

\end{itemize}

\subsection{PjRL plot in Projective Real Plane}

\begin{figure} 
\begin{center}
\includegraphics[scale=1.4]{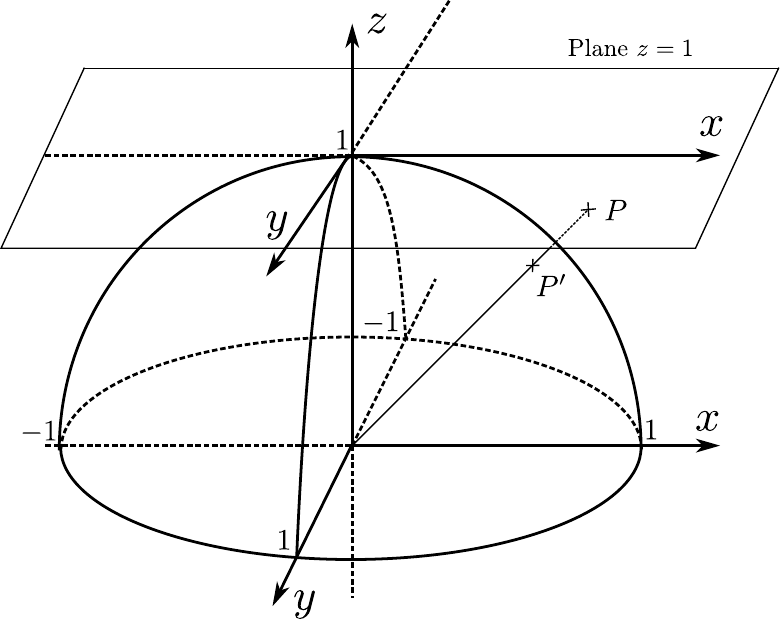}
\caption{\label{hsphere}Gnomonic Projection of half unit sphere onto plane $z=1$}
\end{center}
\end{figure}

In order to plot the PjRL we can use the concept of gnomonic projection 
(\cite{coxeter}) to obtain a geometric representation the 
projective real plane. In this representation, $\mathbb{P}^2(\mathbb{R})$ 
is identified with a semi-sphere of radius one in $\mathbb{R}^3$, as shown 
in Figure~\ref{hsphere}. We note that the points $P$ (on the plane) and $P'$ 
(on the sphere surface) in 
Figure~\ref{hsphere} have the same homogeneous coordinates, since 
they belong to the same line in $\mathbb{R}^3$. The points at infinity 
in $\mathbb{P}^2(\mathbb{R})$ are 
identified with the equatorial great circle, remembering that antipodal 
points (opposite relative to the sphere center) 
have the same homogeneous coordinates.
Also, we note that any line in the plane $z=1$ 
corresponds to a great semicircle on the 
semi-sphere and the the left (right) $z$-semi-plane
corresponds to the left (right) half of the 
semi-sphere. So, the PjRL plot is made onto this semi-sphere, and its 
gnomonic projection onto the plane $z=1$ coincides with the 
conventional RL (see examples in Section~\ref{examples}).

\subsection{Complementary Root-Locus plot in $\pmb{ZY}$ affine plane}

The equations for the PjRL will reduce to the equations for the RL 
when we set $z=1$ in the set of homogeneous
polynomials $\{g_{1}^{h}, \ldots g_{s}^{h}\}$ that defines ${\cal W}$. 
This means that when this set of polynomials is de-homogenized with respect the 
variable $z$ we obtain the the RL, that is the intersection of the PjRL
with the affine $XY$ plane. But since the projective plane contains
three sets that are copies of the affine planes $XY$, $ZY$ and $XZ$, 
the PjRL also can give another view of the RL plot, 
when we analyze the intersection of it with the affine plane $ZY$, 
for instance. In 
this situation we de-homogenize the set of polynomials
$\{g_{1}^{h}, \ldots g_{s}^{h}\}$ with respect to the variable $x$, instead of $z$, and obtain
a set of polynomials that defines a new affine variety in $ZY$ plane 
that we will denominate it ``complementary RL''. So, the complementary RL can been as a
gnomonic projection onto the plane $x=1$, instead of onto the plane $z=1$ as shown in 
Figure~\ref{hsphere}. Geometrically, the switch 
of the role of variables $x$ and $z$ in the complementary RL have the effect of ``moving'' 
all the points over the line $x=0$ (in $XY$ plane) to the infinite
and ``bringing'' the points at infinity ($z=0$) to a finite position.
Intuitively we could state that the conventional RL is a plot as seen from the 
beginning ($k=0$) while the complementary RL is a plot as seen from the end 
($k=\infty$). Also there exists an interesting relation between $Y$ crossing
points in the conventional RL and asymptotes in complementary RL. More specifically, 
if the RL crosses the $Y$ axis at a point, say,
$(0:y:1)$ in $XY$ plane for a given value of $k$, when we translate 
this point to the 
$ZY$ plane it will become $(1:y:0)$, that is 
a point a infinity, in fact an asymptote with 
slope $y/1=y$ in $ZY$ plane. So, we conclude that the $Y$ axis crossing points
by the RL will become asymptotes in  complementary RL, and the absolute
value of the variable $z$ will explode to infinity for the corresponding
value of $k$. We also can make a similar analysis, now considering 
the intersection of the PjRL with the affine plane $XZ$.
In examples presented in Section~\ref{examples} we explore the concept of 
complementary RL with concrete computations.

\section{Examples}\label{examples}

As a matter of fixing ideas, we present a series of examples below.

\begin{example}\label{ex3}\em
Let be $G(s) = s/(s^2+1)$. In this case, using notation introduced in 
Equation~(\ref{qdrd}), we easily obtain:
\[
q_d = x^2-y^2+1, \quad r_d = 2xy, \quad q_n = x \quad r_n = y
\]
and using the definition of $q$ and $r$ in (\ref{poly1}--\ref{poly2}), we
have:
\begin{equation}\label{qrex3}
q(x,y,k_d,k_n) = k_d(x^2-y^2+1)+k_nx, \quad r(x,y,k_d,k_n) = k_d(2xy)+k_ny
\end{equation}
Now we compute the Grobner basis for the ideal $\langle q,r \rangle$
using the graded reversed lexicographic order (\cite[pp.~56]{clo}), with 
$x>y>k_d>k_n$. 
We used the software Macaulay2 (\cite{mac})
to make the computations and obtained the Grobner basis
$\{g_1,g_2,g_3,g_4\}$, where:
\begin{eqnarray}
g_{1}(x,y,k_d,k_n) & = & 2xyk_d+yk_n \quad (=r)\label{gb1e3}\\
g_{2}(x,y,k_d,k_n) & = & x^2k_d -y^2k_d + xk_n + k_d \quad (=q)\label{gb2e3}\\
g_{3}(x,y,k_d,k_n) & = & x^2yk_n + y^3k_n-yk_n\label{gb3e3}\\
g_{4}(x,y,k_d,k_n) & = & 2y^3k_d - xyk_n-2yk_d\label{gb4e3}
\end{eqnarray}
and the homogenized polynomials $g_{i}^{h}$ of the Grobner 
basis are obtained using the
procedure indicated in the Introduction:\footnote{in fact, since the polynomials
$g_i$ are already homogeneous relative to $k_d$ and $k_n$, we can homogenize
them relative only to $x$ and $y$, and the resulting $g_{i}^{h}$ will be the same. For 
example, $g_{1}$ could be homogenized as $g_{1}^{h}=z^2g_1(x/z,y/z,k_d,k_n)$.}

\begin{eqnarray}
g_{1}^{h} & = & z^3g_1(x/z,y/z,k_d/z,k_n/z) = 2xyk_d+yzk_n \label{gh1e3}\\
g_{2}^{h} & = & z^3g_2(x/z,y/z,k_d/z,k_n/z) = x^2k_d -y^2k_d + xzk_n + z^2k_d \label{gh2e3}\\
g_{3}^{h} & = & z^4g_3(x/z,y/z,k_d/z,k_n/z) = x^2yk_n + y^3k_n-yz^2k_n \label{gh3e3}\\
g_{4}^{h} & = & z^4g_4(x/z,y/z,k_d/z,k_n/z) = 2y^3k_d - xyzk_n-2yz^2k_d \label{gh4e3}
\end{eqnarray}
The PjRL is the projection onto ${\mathbb P}^2({\mathbb R})$ of
the projective algebraic variety ${\cal W}$ defined by the 
four polynomials $g_{i}^{h}$ presented in 
Eqs.~(\ref{gh1e3}--\ref{gh4e3}) above. We will represent 
this projection by ${\cal W}_k$, where $k\in\mathbb P^1(\mathbb R)$.

\begin{itemize}
\item Initial points of the PjRL (${\cal W}_0)$: Setting $k_n=0$ and
$k_d=1$ in Eqs.~(\ref{gh1e3}--\ref{gh4e3}), we obtain:
$g_{1}^{h} = 2xy$, $g_{2}^{h} = x^2-y^2+z^2$, $g_{3}^{h} = 0$ and $g_{4}^{h} =
2y(y^2-z^2)$, and we see that $g_{1}^{h}=0$ implies $x=0$ or $y=0$. If 
$x=0$, by $g_{2}^{h}=0$ we obtain $y^2=z^2$. In this case we cannot have
$y=0$ or $z=0$, since $(0,0,0)$ is not valid as a solution. So 
we can set $z=1$ ($XY$ plane) and obtain $y=\pm 1$. Then the initial 
points onto plane $z=1$ are
\[
{\cal W}_0 = \{(0:1:1),(0:-1:1)\} = \{(0,1),(0,-1)\}. 
\]
Or, onto the semi-sphere of radius one:
\[
{\cal W}_0 = \left\{\left(0,1/\sqrt{2},1/\sqrt{2}\right),\left(0,-1/\sqrt{2},1/\sqrt{2}\right)\right\}. 
\]

\item Terminal points of the PjRL (${\cal W}_{\infty}$): Setting $k_n=1$
and $k_d=0$ in Eqs.~(\ref{gh1e3}--\ref{gh4e3}), we obtain:
$g_{1}^{h}=yz$, $g_{2}^{h}=xz$, $g_{3}^{h}=y(x^2+y^2-z^2)$ and
$g_{4}^{h}=-xyz$. To solve $g_{i}^{h}=0$, we can simplify the 
set of equations calculating a Grobner basis with this set of polynomials.
The new Grobner basis has three polynomials: $g_{1}^{h}=yz, 
g_{2}^{h}=xz, g_{3}^{h}=y(x^2+y^2)$. Then we have two kinds
of points:
\begin{enumerate}
\item Points at finite position ($z\neq 0$): In this case, solving
$g_{i}^{h}=0$,  we get
$y=0$ and $x=0$. So, the homogeneous coordinate of the 
point is $(0:0:z)=(0:0:1)$ which corresponds to $(0,0)$ onto affine 
plane $z=1$.
\item Points at the infinite plane $H$ ($z=0$): In this case we obtain:
$g_{1}^{h}=0$, $g_{2}^{h}=0$, and $g_{3}^{h}=y(x^2+y^2)$. 
We must have $y=0$ and the unique possible
nonzero solution is
$(x,0,0)$, $x\neq 0$ whose homogeneous coordinate is 
$(x:0:0)=(1:0:0)\}$, which
corresponds to the intersection point of
the {\em horizontal} lines in the plane $XY$ or a pair of points
$(\pm 1,0,0)$ onto the equatorial great circle over the half 
sphere of radius one. We then have: 
\[
{\cal W}_{\infty}=\{(0:0:1),(1:0:0)\}.
\]
\end{enumerate}

\item Intermediary points of the PjRL (${\cal W}_{\lambda}$): In this case we set $k_d=1$  
and $k_n=\lambda$ in the polynomials shown in Eqs.~(\ref{gh1e3}--\ref{gh4e3})
and recalculate the Grobner basis for the resulting set of polynomials to 
obtain:
\begin{equation}\label{intex3}
g_{1}^{h} = y(2x+z\lambda), \quad g_{2}^{h} = x^2 -y^2 + z^2 + xz\lambda, 
\quad g_{3}^{h}=y(x^2+y^2-z^2)
\end{equation}
We see that we must have $z\neq 0$ in (\ref{intex3}), since $z=0$ will 
imply $x=y=0$ which is not valid as a solution; so all intermediary points
are at finite positions.
\end{itemize}

To plot the PjRL over the unit semi-sphere with radius one we need to add the 
equation $x^2+y^2+z^2=1$, with $z\ge 0$ to the set of equations (\ref{intex3}). 
The sketch of the PjRL plot is shown in Figure~\ref{pjrlss2}.
To obtain the conventional RL plot, we set $z=1$ in (\ref{intex3}) and get
$g_{1} = y(2x+\lambda), g_{2}= x^2 -y^2 + x\lambda + 1,$ and  
$g_{3}=y(x^2+y^2-1)$. We note that that if $y\neq 0$, $g_3=0$ will 
require $x^2+y^2=1$. The complete plot for $g_i=0$ is show in Figure~\ref{rl3}.

Now we will analyze the complementary RL (in plane $ZY$): 
Switching the roles of the $x$ and $z$ axis in the 
projective plane, the coordinate $(x:y:z)$ will become $(z:y:x)$. 
Then, re-analyzing the initial and terminal points 
calculated above, we have:
\[
{\cal W}_0^{'} = \{(1:1:0),(1,-1:0) \quad {\rm and} \quad 
{\cal W}_{\infty}^{'} = \{(1:0:0),(0:0:1)\}
\]
We note that the initial points now are at infinity, that 
is they are asymptotes with rates $\pm 1$. Regarding the 
terminal points, there is one at infinity, that is $(1:0:0)$,
or a horizontal asymptote; and other at origin $(0:0:1)$.
To obtain the intermediary points, we set 
$x=1$ in the polynomials shown in (\ref{intex3})
above and we easily see that, if $y\neq 0$, 
we have the hyperbola $z^2-y^2=1$. The plot for both
conventional and complementary RL are shown in Figure~\ref{rl3}.

\begin{figure} 
\begin{center} 
\includegraphics[scale=1.4]{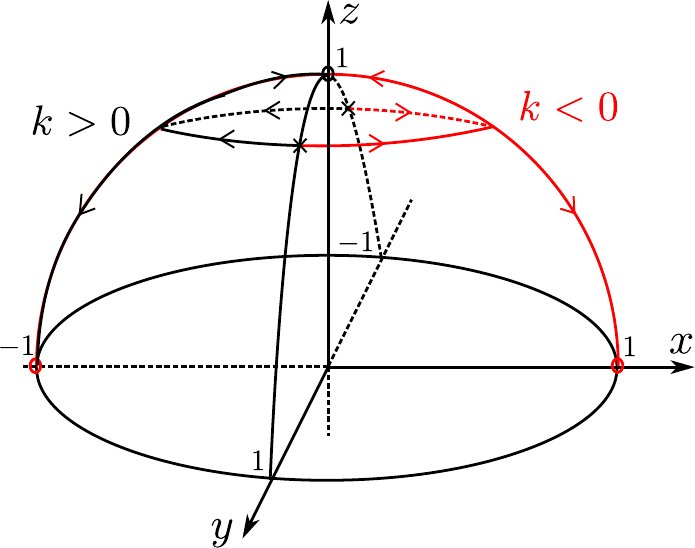}
\caption{\label{pjrlss2} PjRL plot for $G(s)=s/(s^2+1)$}
\end{center}
\end{figure}

\begin{figure} 
\begin{center}
\includegraphics[scale=1.5]{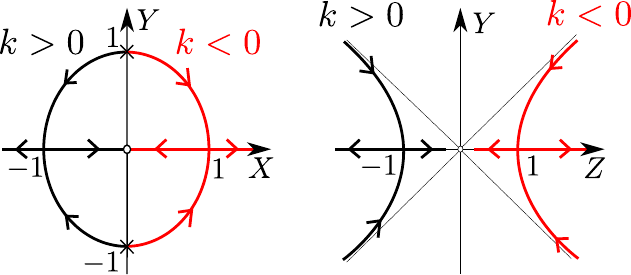}
\caption{\label{rl3}Conventional and complementary RL for $G(s)=s/(s^2+1)$}
\end{center}
\end{figure} 

\begin{remark}\em
We note, in this example, that if we directly homogenize the polynomials $q$ and 
$r$, defined in Equation~(\ref{qrex3}) (instead of the Grobner basis polynomials 
shown in Equations~(\ref{gb1e3}--\ref{gb4e3})),
we are only left with two equations, namely:
\[
q^h = x^2k_d -y^2k_d + xzk_n + z^2k_d \quad {\rm and} \quad
r^h = 2xyk_d+yzk_n,
\]
as opposed to the four equations (\ref{gh1e3}--\ref{gh4e3}). So, if  we 
evaluate the terminal points ${\cal W}_{\infty}$ using only $q^h$ and 
$r^h$ above, we would have, after setting $k_d=0$ and $k_n=1$, 
$q^h=xz$ and $r^h=yz$. It is easy to see, by
these two last equations, that $(1:1:0)$ would be a possible terminal
point at infinity, and this point, as we know, is spurious. This 
confirms the  point discussed above, that we need to evaluate
the Grobner basis for the ideal $\langle q, r\rangle$.

\end{remark}

\end{example}

\begin{example}\label{ex1}\em
Let be $G(s) = (s+1)/s^2$, so we have
\[
q_d(x,y) = x^2-y^2, \quad r_d(x,y)=2xy, \quad q_n(x,y)=x+1, \quad r_n(x,y)=y
\]
and using the definition of $q$ and $r$ in (\ref{poly1}--\ref{poly2}), we
have:
\[
q(x,y,k_d,k_n) = k_d (x^2-y^2)+k_n(x+1), \quad
r(x,y,k_d,k_n) = 2k_d xy + k_n y.
\]
Now we compute the Grobner basis for the ideal $\langle q,r \rangle$
using the graded reversed lexicographic order with 
$x>y>k_d>k_n$ and obtain $\{g_1,g_2,g_3,g_4\}$, where:
\begin{eqnarray*}
g_1(x,y,k_d,k_n) & = & 2xyk_d+yk_n \quad (= r) \\
g_2(x,y,k_d,k_n)  & = & x^2k_d-y^2k_d+xk_n+k_n \quad (= q)\\
g_3(x,y,k_d,k_n)  & = & x^2yk_n+y^3k_n+2xyk_n \\
g_4(x,y,k_d,k_n)  & =& 2y^3k_d-xyk_n-2yk_n
\end{eqnarray*}
Now we homogenize of the polynomials $g_i$, using the
procedure indicated in the Introduction:
\begin{eqnarray}
g_{1}^{h} & = & z^3g_1(x/z,y/z,k_d/z,k_n/z) = 2xyk_d + yzk_n \label{gh1e1}\\
g_{2}^{h} & = & z^3g_2(x/z,y/z,k_d/z,k_n/z) = x^2k_d-y^2k_d+xzk_n+z^2k_n \label{gh2e1}\\
g_{3}^{h} & = & z^4g_3(x/z,y/z,k_d/z,k_n/z) = x^2yk_n + y^3k_n + 2xyzk_n \label{gh3e1}\\
g_{4}^{h} & = & z^4g_4(x/z,y/z,k_d/z,k_n/z) = 2y^3k_d - xyzk_n - 2yz^2k_n \label{gh4e1}
\end{eqnarray}
The PjRL is the set of projective algebraic varieties ${\cal W}_k$
generated by the four polynomials
$g_{i}^{h}$ presented in Eqs.~(\ref{gh1e1}--\ref{gh4e1}) above, 
for each $k=k_n/k_d\in \mathbb P^1(\mathbb R)$.

\begin{itemize}
\item Initial points of the PjRL (${\cal W}_0)$: Setting $k_n=0$ and
$k_d=1$ in Eqs.~(\ref{gh1e1}--\ref{gh4e1}), we obtain:
$g_{1}^{h} = 2xy$, $g_{2}^{h} = x^2-y^2$, $g_{3}^{h} = 0$ and $g_{4}^{h} =
2y^3$, and we see that the
simultaneous solution for $g_{i}^{h}=0$ 
is $(0,0,z)$ where $z\in\mathbb R$. The homogeneous
coordinate for this point
is $(0:0:z)$, but since $(0:0:0)$ is undefined we need $z\neq 0$ and the
``unique'' homogeneous coordinate
possible is $(0:0:1)$. Then we have:
\[
{\cal W}_0=\{(0:0:1)\},
\]
which represents the point
$(0,0)$ in the affine plane ($XY$).
\item Terminal points of the PjRL (${\cal W}_{\infty}$): Setting $k_n=1$
and $k_d=0$ in Eqs.~(\ref{gh1e1}--\ref{gh4e1}), we obtain:
$g_{1}^{h}=yz$, $g_{2}^{h}=xz+z^2$, $g_{3}^{h}=x^2y+y^3+2xyz$ and
$g_{4}^{h}=-xyz-2yz^2$. To solve $g_{i}^{h}=0$, we can simplify the 
set of equations by recalculating a new Grobner basis with this set of polynomials.
The new Grobner basis has three polynomials: $g_{1}^{h}=yz, 
g_{2}^{h}=xz+z^2, g_{3}^{h}=x^2y+y^3$. Then we have two kinds
of points:
\begin{enumerate}
\item Points at the affine plane $XY$ ($z=1$): In this case we get:
$g_{1}= y$, $g_{2}=x+1$, and 
$g_{3}= y(x^2+y^2)$. We see that the unique
solution possible
is $(-1,0,1)$ whose homogeneous coordinate is $(-1:0:1)$, and
this (homogeneous) point corresponds to
$(-1,0)$ in the affine plane $XY$.
\item Points at the infinite plane $H$ ($z=0$): In this case we obtain:
$g_{1}=0$, $g_{2}=0$, and
$g_{3}=x^2y+y^3=y(x^2+y^2)$. We must have $y=0$ and the unique possible
nonzero solution is
$(x,0,0)$, $x\neq 0$ whose homogeneous coordinate is 
$(x:0:0)=(1:0:0)\}$, which
corresponds to the intersection point of
the {\em horizontal} lines in the plane $XY$. By the RL plot for this 
$G(s)$ (Figure~\ref{drl1}) we 
see that this is the direction of the asymptotes when $k\to\pm\infty$.
\end{enumerate}
We then have: 
\[
{\cal W}_{\infty}=\{(-1:0:1),(1:0:0)\}.
\]

\item Intermediary points of the  PjRL (${\cal W}_{\lambda}$): Setting
$k_n=\lambda\neq 0$ and $k_d=1$ in Eqs.~(\ref{gh1e1}--\ref{gh4e1})
we obtain: $g_{1}^{h}=2xy+yz\lambda$,
$g_{2}^{h}=x^2-y^2+xz\lambda+z^2\lambda$, $g_{3}^{h}=\lambda(x^2y+y^3+2xyz$),
$g_{4}^{h}=2y^3-xyz\lambda-2yz^2\lambda$. Computing a new Grobner basis 
with this set of polynomials we obtain three polynomials:
\begin{equation}\label{wl} 
g_{1}^{h}=2xy+yz\lambda, \quad g_{2}^{h}=x^2-y^2+xz\lambda+z^2\lambda, 
\quad g_{3}^{h}=x^2y+y^3+2xyz
\end{equation}

We easily see that $z=0$ will
imply $x=y=0$ and, since we are interested in nonzero
solutions, we must have $z\neq 0$ (implying that doesn't exists intermediary
points in the infinite plane $H$). To see the graph in the $XY$ plane, we set 
$z=1$ in these polynomials
and obtain $g_{1}=y(2x+\lambda)$, $g_{2}=x^2-y^2+\lambda(x+1)$,
and $g_{3}=y(x^2+y^2+2x)=y[(x+1)^2+y^2-1]$.
It is easy to see that solving $g_{i}=0$ we obtain the same set of
affine varieties ${\cal V}_{\lambda}$, $\lambda\in\mathbb R$,
that represents the RL for $G(s)$ in the affine $XY$ plane, as shown 
in Figure~\ref{drl1}.
\end{itemize}

We now analyze the question of obtaining the ``complementary RL'', that 
corresponds to the intersection of the PjRL with $ZY$ plane. 
We note that switching the roles of the $x$ and $z$ axis in the 
projective plane, the coordinate $(x:y:z)$ will become $(z:y:x)$. 
Then, re-analyzing the points calculated above, we have:
 
\begin{itemize}

\item Initial points: The projective point ${\cal W}_0 = \{(0:0:1)\}$ will become 
${\cal W}_{0}^{'} = \{(1:0:0)\}$ and this corresponds to a point at infinity in the 
affine plane $ZY$ (intersection of all horizontal lines).

\item Terminal points: The projective points ${\cal W}_{\infty} = \{(-1:0:1),(1:0:0)\}$
will become ${\cal W}_{\infty}^{'} = \{(1:0:-1)=(-1:0:1), (0:0:1)\}$. This corresponds to the 
points $(-1,0)$ and $(0,0)$ in the affine plane $ZY$.

\item Intermediary Points: To obtain the intermediary points ${\cal W}^{'}_{\lambda}$
in the affine plane $ZY$ we set $x=1$ in (\ref{wl}) to obtain:
\begin{eqnarray*}
g_1(y,z) & = & 2y + yz\lambda = y(2+z\lambda) \\
g_2(y,z) & = & 1-y^2 + z\lambda + z^2\lambda \\
g_3(y,z) & = & y + y^3 + 2yz = y(1+y^2+2z)
\end{eqnarray*}
We only have two cases ($y=0$ and $y\neq 0$):
\begin{enumerate}
\item $y=0$: In this case we are left only with $g_2(0,z)=0$ or $\lambda z^2 + \lambda z + 1=0$ or
$z^2 + z +1/\lambda=0$, whose roots are $z_{1,2}=(-1\pm\sqrt{1-4/\lambda})/2$.
\item $y\neq 0$: In this case, from $g_3(y,z)=0$ we have $y^2+2z+1=0$,
which is the parabola $z=-y^2/2-1/2$.
\end{enumerate}
\end{itemize}
The plot for this set of equations in plane $ZY$ is shown in Figure~\ref{drl1}.

\begin{figure} 
\begin{center}
\includegraphics[scale=1.4]{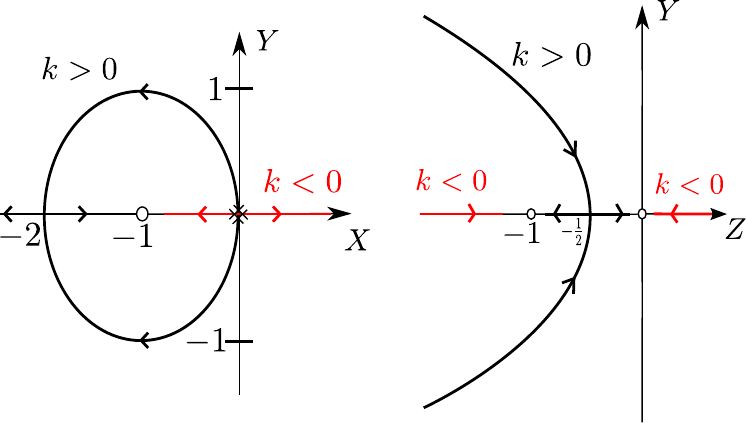}
\caption{\label{drl1}Conventional and complementary RL for $G(s)=(s+1)/s^2$}
\end{center}
\end{figure} 

\end{example}

\begin{example}\label{ex4}\em
Let be $\displaystyle G(s)=\frac{1}{s((s+4)^2+4^2)}$. In this case we have:
\[
q = k_d(x^3-3xy^2+8x^2-8y^2+32x)+k_n, \quad r = k_d(-y^3+3yx^2+16xy+32y)
\]
and computing the Grobner basis using the 
graded reversed lexicographic order with 
$x>y>z>k_d>k_n$, we obtain the following set of
homogenized Grobner polynomials:

\begin{eqnarray}
g_{1}^{h} & = & 3x^2yk_n -y^3k_n + 16xyzk_n + 32yz^2k_n \label{gh1e4}\\
g_{2}^{h} & = & 3x^2yk_d -y^3k_d + 16xyzk_d + 32yz^2k_d \label{gh2e2}\\
g_{3}^{h} & = & x^3k_d - 3xy^2k_d + 8x^2zk_d -8y^2zk_d + 32 xz^2k_d + k_nz^3 \label{gh3e4}\\
g_{4}^{h} & = & 24xy^3k_d + 64y^3zk_d - 64 xyz^2k_d + 256yz^3k_d - 9yz^3k_n\label{gh4e4}\\
g_{5}^{h} & = & 24y^5k_d - 320y^3z^2k_d + 1280xyz^3k_d - 27 xyz^3k_n + 4096yz^4k_d - 72yz^4k_n\label{gh5e4}
\end{eqnarray}

\begin{itemize}
\item Initial points (${\cal W}_0$): In spite of knowing that the 
initial points are the roots of $d(s)$ we calculate them here just as a matter of checking
the theory. Setting $k_d=1$ and $k_n=0$ in the  polynomials above we get:

\begin{eqnarray*}
g_{1}^{h} & = & 0\\
g_{2}^{h} & = & 3x^2y -y^3+ 16xyz + 32yz^2 \\
g_{3}^{h} & = & x^3 - 3xy^2 + 8x^2z -8y^2z+ 32 xz^2 \\
g_{4}^{h} & = & 24xy^3 + 64y^3z - 64 xyz^2 + 256yz^3\\
g_{5}^{h} & = & 24y^5 - 320y^3z^2 + 1280xyz^3 + 4096yz^4 
\end{eqnarray*}
We note that doesn't exists initial points at infinity, since if we choose $z=0$, by $g_{5}^{h}=0$ we necessarily have
$y=0$ and, by $g_{3}^{h}=0$ we get $x=0$, but $(0:0:0)$ is not valid as solution. So, all the initial points are onto affine
$XY$ plane; then setting $z=1$ in the homogeneous polynomials above we obtain the set of (non-homogeneous)
polynomials:
\begin{eqnarray*}
g_{2} & = & 3x^2y -y^3+ 16xy + 32y = y(3x^2-y^2+ 16x + 32) \\
g_{3} & = & x^3 - 3xy^2 + 8x^2 -8y^2+ 32 x \\
g_{4} & = & 24xy^3 + 64y^3 - 64 xy + 256y = y(24xy^2 + 64y^2 - 64 x + 256)\\
g_{5} & = & 24y^5 - 320y^3 + 1280xy + 4096y = y(24y^4 - 320y^2 + 1280x + 4096)
\end{eqnarray*}
We have two cases, namely $y=0$ or $y\neq 0$; if we set $y=0$ the only possible real value for $x$ is 
obtained from $g_{3}=0$ and it is $x=0$. So the first initial point is $(0:0:1)$. Considering $y\neq 0$ in the
polynomials above we can rewrite them as:
\begin{eqnarray*}
g_{2}/y & = & 3x^2 -y^2+ 16x + 32 \\
g_{3} & = & x^3 - 3xy^2 + 8x^2 -8y^2+ 32 x \\
g_{4}/y & = & 24xy^2 + 64y^2 - 64 x + 256\\
g_{5}/y & = & 24y^4 - 320y^2 + 1280x + 4096
\end{eqnarray*}
and if we compute a Grobner basis for the set of polynomials above we obtain:
\begin{eqnarray*}
h_1 & = & 3x^2 -y^2+ 16x + 32 \\
h_3 & = & 3xy^2 + 8y^2 - 8x + 32\\
h_4 & = & 3y^4 - 40y^2 + 160x + 512
\end{eqnarray*}
To solve the set of equations $h_1=h_2=h_3=0$, we can eliminate $y^2$ from $h_1$ 
and $h_2$ to obtain the equation $x^3+8x^2+24x+32=0$ whose only real solution 
is $x=-4$, and this implies $y=\pm 4$. Finally we have the following set of initial points:
\[
{\cal W}_0 = \{(0:0:1),(-4:4:1),(-4:-4:1)\}
\]

\item  Terminal points (${\cal W}_{\infty}$): Setting $k_d=0$ and $k_n=1$ in Eqs.~(\ref{gh1e4}--\ref{gh5e4})
we obtain:
\begin{enumerate}
\item Points onto affine plane $XY$ ($z=1$): This will make $g_{3}^{h}=1$, what means that doesn't 
exist terminal points onto affine plane $XY$.

\item Points at infinity ($z=0$): We are left, from (\ref{gh1e4}), only with $3x^2y-y^3=y(3x^2-y^2) = 0$.
If $y=0$, we will have the solution $(x:0:0)=(1:0:0)$; if $y\neq 0$, we will have $y^2=3x^2$, that results 
in points $(1:\pm\sqrt{3}:0)$. So, the terminal points are:
\[
{\cal W}_{\infty} = \{(1:0:0),(1:\sqrt{3}:0),(1:-\sqrt{3}:0)\}
\]
\end{enumerate}
\item Intermediary points (${\cal W}_{\lambda}$): Setting $k_d=1$, $k_n=\lambda\neq 0$ in Eqs.~(\ref{gh1e4}--\ref{gh5e4}),
and computing a new Grobner basis we have the set below with only two polynomials:
\begin{equation}\label{intex4}
g_{1}^{h} = 3x^2y-y^3+16xyz+32yz^2, \quad {\rm and} \quad 
g_{2}^{h} = z^3\lambda +x^3 - 3xy^2+8x^2z-8y^2z+32xz^2
\end{equation}

\end{itemize}

In this example we will only analyze the complementary RL (in plane $ZY$), 
and for that we will recalculate the initial and terminal points, just by 
switching the position of $x$ and $z$ in ${\cal W}_0$ and ${\cal W}_{\infty}$. So the new initial and terminal 
points are:
\[
{\cal W}_0^{'} = \{(1:0:0),(-1/4:-1:1),(-1/4:1:1)\}
\]
and
\[
{\cal W}_{\infty}^{'} = \{(0:0:1),(0:\sqrt{3}:1),(0:-\sqrt{3}:1)\}.
\]
To evaluate the intermediary points we set $x=1$ in the polynomials shown in
(\ref{intex4}) to obtain:
\begin{eqnarray*}
h_1 & = & 3y-y^3+16yz+32yz^2= y(3-y^2+16z+32z^2) \\
h_2 & = & z^3\lambda - 8y^2z - 3y^2 + 32z^2 + 8z + 1
\end{eqnarray*}
And clearly we have two cases to consider, namely $y=0$ and $y\neq 0$:
\begin{itemize}
\item $y=0$ will imply $h_1=0$ and $h_2=z^3\lambda+32z^2+8z+1$. We easily see
that the cubic polynomial $h_2$ always have one real and two complexes roots. The real
root varies with $\lambda$ as shown in Figure~\ref{rl4}.
\item $y\neq 0$ will imply $(-h_1/y) = y^2-32z-16z-3$, so we have the polynomials:
\begin{eqnarray*}
-h_1/y & = & y^2-32z^2-16z-3\\
h_2 & = & z^3\lambda - 8y^2z - 3y^2 + 32z^2 + 8z + 1
\end{eqnarray*}
Now if we compute again a Grobner basis for this set we will have:
\begin{eqnarray*}
l_1 & = & y^2-32z^2-16z-3\\
l_2 & = & (\lambda - 256)z^3 - 192z^2 - 64z - 8
\end{eqnarray*}
We note that $l_1=0$ represents the hyperbole shown in Figure~\ref{rl4}, while
$l_2=0$ determines as $z$ depends on $\lambda$. We note that when 
$\lambda\neq 256$, $l_2$ is a cubic polynomial with one real and two complex
roots; moreover the real root explodes to infinity when $\lambda\to 256$. We 
also note that 256 is the value of the gain $k$ where the RL crosses the $Y$ 
axis (in plane $XY$).

\end{itemize}

\begin{figure} 
\begin{center}
\includegraphics[scale=2]{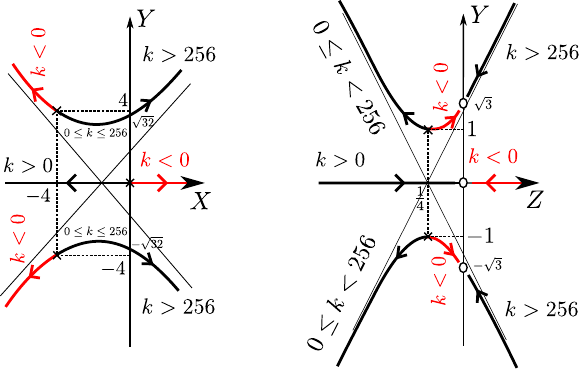}
\caption{\label{rl4}Conventional and complementary RL for $\displaystyle G(s)=\frac{1}{s((s+4)^2+4)}$}
\end{center}
\end{figure} 

\end{example}

\begin{example}\label{ex5}\em

We now consider a case where the degree of $d$ is equal the degree of $n$. 
The main point here is that the polynomial $d(s)+kn(s)$ may decrease its degree
for some (finite) value of $k$. To simplify the PjRL plot we will consider a simple 
rational function $G$ defined
as:
\[
G(s) =\frac{1-s^2}{1+s^2}
\]
In this example we easily see that $(1+s^2)+(1-s^2)=2$, so the degree of 
$d(s)+kn(s)$ is zero for $k=1$, and we have no finite roots.
To analyze the PjRL we evaluate 
the polynomials $q$ and $r$, defined in (\ref{poly1}--\ref{poly2}), which in this case are:
\[
q= k_d(1+x^2-y^2) + k_n(1-x^2+y^2), \quad {\rm and} \quad
r= k_d(2xy) + k_n(-2xy)
\]

Computing the Grobner basis using the 
graded reversed lexicographic order with 
$x>y>z>k_d>k_n$, we obtain the following set of
homogenized Grobner polynomials:
\begin{eqnarray}
g_{1}^{h} & = & xyk_n \label{gh1e5}\\
g_{2}^{h} & = & xyk_d \label{gh2e5}\\
g_{3}^{h} & = & k_d(z^2+x^2-y^2) + k_n(z^2+y^2-x^2) \label{gh3e5}\\
g_{4}^{h} & = & y(y^2k_d-y^2k_n-z^2k_d-z^2k_n)\label{gh4e5}
\end{eqnarray}

\begin{itemize}
\item Initial points (${\cal W}_0$): Using $k_d=1$ and $k_n=0$ in 
equations (\ref{gh1e5}--\ref{gh4e5}) above we obtain:
\[
g^{h}_{2}=xy, \quad g^{h}_{3}=z^2+x^2-y^2, \quad {\rm and} \quad 
g^{h}_{4} = y(y^2-z^2)
\]
and the unique (non-null) possible solution is $x=0$, $y\neq 0$ and 
$y^2=z^2$; so the homogeneous coordinates for ${\cal W}_0$
is:
\[
{\cal W}_0 = \{(0:1:1), (0:-1:1)\}
\]
On the semi-sphere of radius one, we have:
\[
{\cal W}_0 = \{(0,1/\sqrt{2},1/\sqrt{2}), (0,-1/\sqrt{2},1/\sqrt{2})\}
\]

\item Terminal points (${\cal W}_{\infty}$): Using $k_d=0$ and $k_n=1$ in 
equations (\ref{gh1e5}--\ref{gh4e5}) above we obtain:
\[
g^{h}_{1}=xy, \quad g^{h}_{3}=z^2-x^2+y^2, \quad {\rm and} \quad 
g^{h}_{4} = -y(y^2+z^2)
\]
and the unique (non-null) possible solution is $x\neq 0$, $y=0$ and 
$x^2=z^2$; so the homogeneous coordinates for ${\cal W}_{\infty}$
is:
\[
{\cal W}_{\infty} = \{(1:0:1), (-1:0:1)\}
\]
On the semi-sphere of radius one, we have:
\[
{\cal W}_{\infty} = \{(1/\sqrt{2},0,1/\sqrt{2}), (-1/\sqrt{2},0,1/\sqrt{2})\}
\]

\item Intermediary points (${\cal W}_{\lambda}$):Using $k_d=1$ and $k_n=\lambda$ in 
equations (\ref{gh1e5}--\ref{gh4e5}) and recalculating the Grobner basis we get:
\begin{eqnarray}
g_{1}^{h} & = & xy \label{gh1me5}\\
g_{2}^{h} & = & (\lambda-1)x^2 + (1-\lambda)y^2 - (1+\lambda)z^2\label{gh2me5}\\
g_{3}^{h} & = & y[-(1-\lambda)y^2+(\lambda+1)z^2)]\label{gh3me5}
\end{eqnarray}
We have two cases to consider, namely, $x=0, y\neq 0$ and $x\neq 0, y=0$
(the case $x=0$ and $y=0$ will imply $z=0$ which is not valid as a solution). 
Also, to plot the PjRL over the semi-sphere of radius one we have to consider the 
restriction $x^2+y^2+z^2=1$ with $z\ge 0$. Then we have:
\begin{enumerate}
\item $x=0$, $y\neq 0 $ and $y^2+z^2=1$. Using equations 
(\ref{gh1me5}--\ref{gh3me5}) defined above, we 
necessarily have $(1-\lambda)y^2 - (1+\lambda)z^2=0$, and then we get:
\[
y=\pm\sqrt{\frac{1+\lambda}{2}}, \quad {\rm and} \quad 
z = \sqrt{\frac{1-\lambda}{2}}
\]
So, considering $\lambda>0$, we must have $0<\lambda \le 1$ and for $\lambda=1$ 
we have the point at infinity $(0,1,0)=(0,-1,0)$.

\item $x\neq 0$, $y=0$ and $x^2+z^2=1$. Using equations 
(\ref{gh1me5}--\ref{gh3me5}) defined above, we 
necessarily have $(\lambda-1)x^2 - (1+\lambda)z^2=0$, and then we get:
\[
x=\pm\sqrt{\frac{\lambda-1}{2\lambda}}, \quad {\rm and} \quad 
z = \sqrt{\frac{\lambda-1}{2\lambda}}
\]
Again, considering $\lambda>0$, we must have $\lambda \ge 1$, and for 
$\lambda=1$ we get the point at infinity $(1,0,0)=(-1,0,0)$.

\end{enumerate}

\end{itemize}

In Figure~\ref{pjrlex5} we show the plot for the PjRL obtained in this example; 
as we can note, the PjRL has a discontinuity at infinity.

\begin{figure} 
\begin{center} 
\includegraphics[scale=1.4]{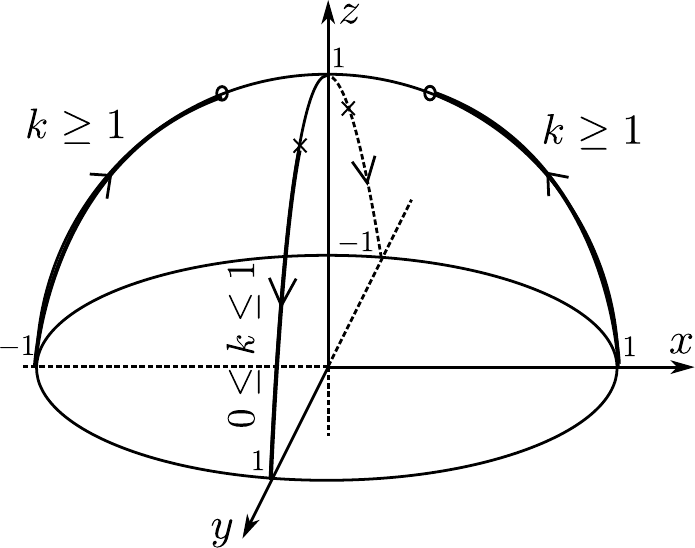}
\caption{\label{pjrlex5} PjRL plot for $G(s)=\displaystyle\frac{1-s^2}{1+s^2}$}
\end{center}
\end{figure} 

\end{example}

\section{Conclusions}
We have presented in this paper an extension of the classical 
Root-Locus (RL) method, denominated Projective Root-Locus (PjRL), where 
the coordinates of the points of the RL for an irreducible rational
function $G(s)=n(s)/d(s)$ are represented in the projective real plane 
$\mathbb P^2(\mathbb R)$ instead of the affine plane $\mathbb R^2$. To
obtain the PjRL we used results from algebraic geometry, representing the RL 
as an affine algebraic variety and extrapolating it to the projective plane. With this 
approach we could obtain the RL points at infinity as
solutions of a set of algebraic equations. Also, we have shown how to plot
the PjRL onto a semi-sphere of radius one that is a representation of 
the projective plane $\mathbb P^2(\mathbb R)$.
Since the real projective plane contains 
three ``copies" of real affine planes, we can plot the RL onto an affine real plane, 
other than the original $XY$ one; we denominated this new plot  ``complementary RL'', and 
we have shown that the points where the RL crosses the $Y$ in $XY$ plane turns into
asymptotes of the complementary RL in $ZY$ affine plane, and vice-versa. Several examples were
worked out in order to show that the PjRL can be relatively 
easily obtained using a computer algebra software.

\bibliographystyle{plain}

\begin{thebibliography}{00}



\bibitem{dh} J. D'Azzo and C. Houpis. {\em Linear Control System Analysis and Design}. Second Edition. 
MacGraw-Hill Kogakusha, Ltd., 1981. 


\bibitem{scilab} Scilab Enterprises. Scilab: Free and Open Source Software for Numerical Computation. Orsay, France, 2012.
Available at http://www.scilab.org. 

\bibitem{clo} D. Cox, J. Litlle and D. O'Shea.{\em Ideals, Varieties, and Algorithms: An Introduction to Computational Algebraic
Geometry and Commutative Algebra}. Second Edition. Springer-Verlag New York Inc., 1997.

\bibitem{sha} I. Shafarevich. {\em Basic Algebraic Geometry 1: Varieties in Projective Space}. Second Edition. Springer-Verlag
Berlin Heildelberg, 1994. 

\bibitem{coxeter} H. Coxeter and S. Greitzer. {\em Geometry Revisited}. Mathematical Association of America, 1967.

\bibitem{mac} D. Grayson and M. Stillman. Macaulay2, A Software System for Research in Algebraic Geometry. Available at 
http://www.math.uiuc.edu/Macaulay2. 


\end{thebibliography}

\end{document}